# Resolving Thermospheric Vertical Wind Ambiguities and Energy Processes

Jeffrey P. Thayer[1,2] & Austin Coleman[1]

[1]University of Colorado, Aerospace Engineering Sciences Department, Boulder CO

[2]University of Colorado, Space Weather Technology, Research, and Education Center, Boulder CO

Corresponding author: Jeffrey P. Thayer (jeffrey.thayer@colorado.edu)

Key Points:

- Vertical winds require a coordinate definition to properly interpret their impact on energy processes in the thermosphere.

- Non-energetic lifting winds can be an appreciable component of upper thermosphere height-based vertical winds on constant pressure surfaces.

- Measures of vertical winds using airglow emissions are susceptible to undetermined lifting wind components.

## Contents

# Abstract

This study applies a generalized vertical coordinate system approach alongside thermodynamic control volume analysis to explore the nuanced interpretations of energy transfer processes associated with vertical motion in the thermosphere. Using simulations from the TIEGCM V3.0 model, a key finding reveals that transforming vertical winds in height coordinates onto constant pressure surfaces contain a substantial lifting component—an aspect often overlooked in previous research. This distinction is critical for internal energy assessments, as vertical winds defined in pressure coordinates directly contribute to the adiabatic heating/cooling rate while only a portion of the vertical wind in height coordinates contributes to these energy changes. These differences are demonstrated through schematic representations of thermodynamic control volumes that model a column of thermospheric gas undergoing various energy transfer processes. Accurately determining the portion of observed vertical winds that affect internal energy requires clear understanding between height and pressure surfaces – a nontrivial challenge in observational contexts. Furthermore, applying the generalized vertical coordinate framework to airglow emissions in the upper thermosphere uncovers an inherent ambiguity: vertical winds inferred from airglow observations may not align with those defined on pressure or altitude surfaces. This insight suggests that discrepancies in the behavior and magnitude of vertical winds derived from FPI observations of red-line emissions in the upper thermosphere are due to poorly known relations between the emission structure relative to pressure or height surfaces.

# 1  Introduction

The thermosphere is a strongly driven system influenced by terrestrial and extraterrestrial sources of energy (Knipp et al., 2004; Thayer and Semeter, 2004; Liu, 2016). The thermospheric energy budget accounts for energy input processes, like Joule heating, solar EUV absorption, and particle precipitation, and energy transfer processes, like heat conduction and radiation. These energy sources entering and leaving the thermosphere are highly impactful in altering its properties. However, changes in the internal energy of the thermospheric gas must also account for inner workings of the gas, such as expansion and compression. This is essentially dictated by the first law of thermodynamics where work done by, or on, the thermospheric gas in the form of expansion/compression work is a key contributor (e.g., Dickinson et al., 1968). Enthalpy is the gas property that accounts for both internal energy and expansion/compression work. Akmeav and Juang (2008) have explored the use of specific enthalpy as a prognostic thermodynamic variable in atmospheric numerical models. A change in enthalpy can involve internal energy changes by heat transfer processes or by expansion/compression work. This form of work accounts for volumetric and/or pressure changes experienced by the gas that either decrease or increase its internal energy and is intimately tied to vertical motions in the thermosphere. Thus, to determine the internal energy, or temperature, of the gas correctly, vertical motion must be accurately represented. Other work processes can operate on the gas that can lead to changes in the gas bulk kinetic energy, i.e. changes in winds, or potential energy, i.e. changes in height, but these are typically arranged and expressed separately by introducing a mechanical energy equation.

Descriptions of vertical motion require a definition of its vertical coordinate. This is often presumed by the units used to describe vertical motion, i.e. a Cartesian vertical wind in units of m/s uses height as its vertical coordinate. However, that presumption may not be sufficient to interpret the energetics associated with that vertical wind, or how an observation using the same units may not be directly comparable. The effort put forth in this paper is to describe a generalized vertical coordinate system and then apply it to pressure, height, and other scalar functions, to define vertical motion in each of those frames and transformations between these frames. The energy consequence of selecting a particular vertical coordinate will be elucidated by performing internal energy assessments.

There exists in the literature a form of paradox in vertical wind observations where large magnitudes are observed that do not conform to theory (e.g., Smith, 1998 and references therein) or do not compare well with other vertical wind observations (Harding et al., 2017 and references therein). One issue that may obfuscate such outcomes and comparisons is the insufficient description of the vertical coordinate system being used to define an observation or model result. For example, the NCAR TIEGCM uses log-pressure as its vertical coordinate but does compute a "vertical wind" in units of m/s. GITM uses height as its vertical coordinate and computes a vertical wind in units of m/s (Yiğit and Ridley, 2011). These two vertical wind outputs are not the same even though the units may coincide. This will become more obvious in the later sections of this paper. Furthermore, a description of vertical winds in units of m/s on a constant pressure surface is commonly described as composed of a barometric wind component and a divergence wind component (i.e., Rishbeth et al., 1969; Rishbeth et al., 1987; Rishbeth and Müller-Wodarg, 1999 and references therein and subsequently). However, this description neglects horizontal advection of height gradients on a constant pressure surface which, as will be shown in later sections, is revealed when applying the generalized vertical coordinate approach. Furthermore, this component will be shown to contribute significantly to the height-based vertical wind on a constant pressure surface when using NCAR-TIEGCM simulations. Moreover, using a generalized vertical coordinate approach to describe airglow emissions in the upper thermosphere demonstrates an inherent ambiguity between airglow vertical winds and vertical winds determined on pressure or height surfaces. These differences suggest that a root cause for differing behavior/magnitudes of vertical winds derived from FPI observations of red-line emissions in the upper thermosphere are due to poorly known relations between the emission horizontal structure relative to pressure or height surfaces.

The focus of this paper is to disambiguate the description of vertical motion, its transformation in coordinates, and its effect on the internal energy of the upper thermospheric gas. Section 2 provides a general description of the internal energy equation for the thermosphere to provide the context from which vertical motions and energy become coupled. Section 3 elucidates the role of vertical motion in the internal energy equation and describes the transformations required to properly account for expansion/compression work when defining vertical motions in different vertical coordinate systems. Section 4 will apply a thermodynamic control volume analysis to the thermospheric gas to assist in the interpretation of the relations derived in Section 3. Section 5 will exercise the descriptions provided in sections 3 and 4 by analyzing NCAR TIEGCM V3.0 internal energy equation and vertical winds to reveal the energy considerations associated with vertical motions defined in different vertical coordinate systems. Section 6 will use the output of the TIEGCM simulations to illustrate adiabatic heating and cooling in the thermosphere and its impact on temperature and density perturbations. Furthermore, a generalized vertical coordinate system will be described to address the issue of vertical wind observations and subsequent analysis/interpretation. The case of airglow observations of vertical winds from the ground is described in this generalized vertical coordinate system to demonstrate that such observations of vertical motion are not expected to be equal to vertical winds in the pressure or height frame unless special conditions of the airglow emission apply. An appendix is also provided to describe the transformation of vertical motions from pressure to height coordinates.

For this paper, the vertical winds produced by the model will be presented without investigation of cause – essentially a kinematic analysis of the upper thermospheric flow field (e.g., Thayer and Killeen, 1991) to study their influence on the thermospheric gas internal energy. The NCAR-TIEGCM V3.0 will be used to evaluate vertical winds defined with pressure as the vertical, independent coordinate. Vertical winds defined with height as the vertical coordinate will also be determined and contrasted with vertical winds defined in pressure coordinates.

## 2  Internal Energy Equation

The first law of thermodynamics establishes that, for a defined system, any change in energy of that system requires a process to take place in the form of heat transfer or work. This change in energy can be of various forms, such as kinetic energy, potential energy, or internal energy. The rate equation for the first law is expressed generally as,

$$\frac{DE}{Dt} = \dot{Q}_{net} - \dot{W}_{net} \tag{1}$$

where the rate of energy change of a discrete system in units of Watts is caused by the net heat transfer rate ($\dot{Q}_{net}$) and net rate of work done by, or on, the system ($\dot{W}_{net}$). This is considered the engineering form of the first law where the rate of work done by the system is defined to be positive valued. The internal energy equation can be parsed from the first law taking account of the heat transfer and work processes that only alter the internal energy, $U$, of the gas.

$$\frac{DU}{Dt} = \dot{Q} - \dot{W} \tag{2}$$

For the thermosphere, general circulation models account for changes in the internal energy as a means to determine the temperature of the gas. The appropriate internal energy equation for the upper thermosphere, expressed in terms of temperature change, requires several steps and defined relationships. The heat transfer rate accounts for energy transferred into and out of the system due to radiation, conduction, or convection. Work is the other means by which a system's energy can change. The rate of work done by the gas has a special form when considering only expansion / compression work operating by, or on, the system while under constant pressure. Such work applies to a gas undergoing a volumetric change while pressure remains fixed, i.e. an isobaric process. This results in the isobaric expansion / compression work rate expressed as,

$\dot{W} = p \frac{D\forall}{Dt}$, where $\forall$ is the symbol for volume to avoid any confusion with velocity.

If the process is also adiabatic, i.e. $\dot{Q} = 0$, and the relation of specific heat at constant volume, $c_\forall$, to internal energy is used, $mc_\forall = DU/DT$, where m is the mass of the gas, then the first law becomes

$$mc_\forall \frac{DT}{Dt} = -p \frac{D\forall}{Dt} \tag{3}$$

This form of the first law leads to the three polytropic (or Poisson) relations between the gas properties of pressure, temperature, and volume for an isobaric, adiabatic process. Atmospheric potential temperature is a commonly used property to determine how much temperature change a parcel of air would experience if it undergoes an isobaric, adiabatic expansion or compression. These conditions also result in the entropy of the gas to be conserved and often isentropic surfaces are used as a vertical coordinate in lower atmosphere systems [e.g., Johnson, 1989].

An important thermodynamic property that combines internal energy with expansion / compression work is enthalpy, defined as $H = U + p\forall$ in units of Joules. Taking the time derivative of enthalpy, $\frac{DH}{Dt} = \frac{DU}{Dt} + \frac{D(p\forall)}{Dt}$, and using the isobaric expansion/compression work in the rate equation, the first law becomes,

$$\frac{DH}{Dt} = \frac{D(p\forall)}{Dt} + \dot{Q} - p\frac{D\forall}{Dt} \qquad (4)$$

Manipulating terms provides the enthalpic form of the first law,

$$\frac{DH}{Dt} = \forall\frac{Dp}{Dt} + \dot{Q} \qquad (5)$$

where the time rate of change in pressure of the gas and heat transfer into or out of the system can cause an enthalpic change to the system. Using the thermodynamic relation for specific heat at constant pressure, i.e. $mc_p = DH/DT$, the first law of thermodynamics can be expressed in the form of temperature as,

$$mc_p \frac{DT}{Dt} = \forall\frac{Dp}{Dt} + \dot{Q} \qquad (6)$$

Note that $c_p$ is always greater than $c_v$ with their difference equal to the gas constant, if ideal gas behavior is assumed. In the thermosphere, ideal gas behavior is highly accurate due to the low pressures and high temperatures experienced by the gas.

The rate equation of the first law, in units of Kelvin per second, expressed below is attained after dividing through by volume and defining a mean specific heat at constant pressure to account for its sensitivity to the range of temperatures expected in the upper thermosphere.

$$\frac{DT}{Dt} = \frac{1}{\rho \overline{c}_p} \frac{Dp}{Dt} + \frac{\dot{Q}^*}{\rho \overline{c}_p} \tag{7}$$

where $\dot{Q}^*$ is now in units of [W/m$^3$] representing all forms of diabatic heat transfer in the thermosphere. It is this form that is commonly applied by thermospheric general circulation models to solve for temperature. The next section evaluates this form of the first law in relation to vertical winds and their associated reference frames.

## 3  Vertical Wind Transformations

In general, the basic laws of physics apply to discrete bounded systems, like that described in section 2, where a material element of gas is analyzed thermodynamically. For a system in motion, a Lagragian framework describes the behavior of properties within a gas material element for a coordinate system that is following the motion. Thus, the total derivative provided in the energy equation of section 2 is the total, material or Lagrangian derivative. An Eulerian description of motion focuses on fixed points in space and how fluid properties change at those locations over time. Such a description may use rectangular Cartesian coordinates $\boldsymbol{x_C} = (x_1, x_2, x_3)$ consisting of orthogonal plane surfaces or spherical coordinates, $\boldsymbol{x} = (x, y, z)$ where $x$ is longitude, $y$ is latitude, and $z$ is height. Spherical coordinates represent an orthogonal curvilinear coordinate system where coordinate lines and surfaces are curved and whose unit vectors are everywhere perpendicular to the other surfaces and point in the direction of their coordinates increasing value.

The Lagrangian derivative expressed in spherical coordinates is given as,

$$\frac{D}{Dt} = \left(\frac{\partial}{\partial t}\right)_z + \vec{V}_h \cdot \vec{\nabla}_z + w \frac{\partial}{\partial z}, \tag{8}$$

where subscript $z$ indicates the operation is performed on a constant height surface. The first term is referred to as the local or Eulerian derivative of a field property at a fixed height accounting for temporal changes in the field property at the position where the material element is located. Commonly in meteorology, this term is called the property tendency and this will be used in subsequent discussions. The second and third terms are referred to as advective derivatives accounting for the change in a field property due to horizontal and vertical motion of the material element to positions of different field values on a constant height surface. Even if local values do not evolve with time, a field property will change if its material element moves across contours of the field property at a rate dictated by its velocity along the direction of the spatial gradient for that field property. The horizontal wind consists of zonal, $u$, and meridional, $v$, components in the $x$- and $y$- coordinates with unit vectors $\mathbf{i}$ and $\mathbf{j}$, respectively. The height-based vertical wind, $w$, is positive in the $\mathbf{k}$-direction which is orthogonal and right-handed to $\mathbf{i}$ and $\mathbf{j}$. Furthermore, the height-based vertical wind is, by definition, $w=Dz/Dt$.

A mapping from one coordinate system to another is unique and invertible if the Jacobian of the transformation matrix is nonzero for all coordinate values. The elements of the Jacobian transformation matrix contain first-order partial derivatives of the new coordinates relative to the old coordinates. If transforming from Cartesian to spherical coordinates, only the diagonal elements are non-zero due to the systems' orthogonality, and these represent the metric scaling factors used in the spherical coordinate system to account for distortions introduced by each of the curvilinear surfaces.

However, a common coordinate system for an Eulerian description of fluid evolution in the Earth's atmosphere replaces the vertical independent variable height, $z$, in the spherical coordinate system with pressure, $p$, and adds a time variable, $t$, such that $\mathbf{x}_p =( x, y, p, t)$, with $x, y, p$, and $t$ being the independent variable and $z=z(x, y, p, t)$ is a four-dimensional dependent variable. The derivation and benefit of expressing the constitutive equations in pressure coordinates for numerical weather prediction has been discussed over the decades by, for example, Sutcliffe (1947), Eliassen (1949), and Kasahara (1974).

As before, this coordinate transformation is unique and invertible if the Jacobian is nonzero. However, the introduction of pressure sets up a special coordinate case where longitude and latitude surfaces remain orthogonal to each other while the pressure surface can be nonorthogonal to these surfaces. This constitutes a nonorthogonal coordinate system. A non-zero determinant of the Jacobian matrix (now with off-diagonal terms due to nonorthogonality) when transforming from height coordinates to pressure coordinates is $\partial p/\partial z$, see Salby (1996) for more details. Thus, a functional relationship must exist between pressure and altitude to perform the mapping. The hydrostatic law often serves as that functional relationship, however, non-hydrostatic behavior can also be addressed if a functional relationship between pressure and altitude can be determined.

With a nonzero Jacobian, the functional relationship is single valued and monotonic, meaning the mapping is invertible where any scalar value may be expressed or transformed between either coordinate system. Furthermore, partial derivatives with respect to $x$, $y$, or $t$ are different depending on whether they are performed with $p$ or $z$ held constant based on the chain rule of functions. For example, if transforming the total derivative, $D/Dt$, from the $z$-system to the $p$-system, the Lagrangian derivative from equation (8) becomes

$$\frac{D}{Dt} = \left(\frac{\partial}{\partial t}\right)_p + \vec{V}_h \cdot \vec{\nabla}_p + \left[w - \left(\frac{\partial z}{\partial t}\right)_p - \vec{V}_h \cdot \vec{\nabla}_p z\right] \frac{\partial p}{\partial z} \frac{\partial}{\partial p} \qquad (9)$$

Note that this relation still employs **i**, **j**, **k** unit vectors as defined before. The Lagrangian derivative expressed in the Eulerian frame for a coordinate system using pressure as its vertical coordinate is,

$$\frac{D}{Dt} = \left(\frac{\partial}{\partial t}\right)_p + \vec{V}_h \cdot \vec{\nabla}_p + \omega \frac{\partial}{\partial p} \qquad (10)$$

The vertical wind in pressure coordinates, $\omega$, is by definition equal to $Dp/Dt$ and given in units of [Pa/s]. Thus, the relationship between $\omega$ and $w$ can be expressed as,

$$\omega = \frac{\partial p}{\partial z}\left[w - \left(\frac{\partial z}{\partial t}\right)_p - \vec{V}_h \cdot \vec{\nabla}_p z\right] \qquad (11)$$

Although properly defined in pressure coordinates, $\omega$ is not commonly measured and so a transformation is required between $\omega$ and the height-based vertical wind, $w$, in the $z$-system. Given that $\omega$ contributes directly to the energy equation provided in equation (7), this transformation will also serve to determine what portion of the height-based vertical wind in the *z*-system contributes to the internal energy equation.

The relationship between a vertical wind in the *p*-system and a vertical wind in the *z*-system evaluated on a constant pressure surface is scaled by the determinant of the Jacobian. Using the hydrostatic relation between pressure and height, this can be written

$$\omega = -\rho g \left\{ w - \left(\frac{\partial z}{\partial t}\right)_p - \vec{V}_h \cdot \vec{\nabla}_p z \right\} \tag{12}$$

If a lifting wind ($w_L$) is defined to be that component of the height-based vertical wind in the *z*-system associated with the local rate of change in height on a fixed pressure surface and the horizontal advection of height along a fixed pressure surface (as defined by Dickinson and Geisler, 1968) then, from equation (12),

$$\omega = -\rho g \{w - w_L\}. \tag{13}$$

Consequently, a vertical wind measured in height coordinates does not directly constitute the contribution to the expansion / compression of a material element that causes a temperature change. In other words, the lifting wind observed in pressure coordinates is the non-energetic portion of the vertical wind defined in the *z*-system. As will be evaluated later, a divergent vertical wind can be defined as

$$w_D = w - w_L \tag{14}$$

and relate directly to ω, when assuming hydrostatic equilibrium, as

$$\omega = -\rho g w_D \tag{15}$$

This evaluation of height-based vertical winds on a constant pressure surface has led to the common description that a vertical wind is composed of a barometric wind component, i.e. $\left(\partial z/\partial t\right)_p$, and a divergence wind component, $w_D$, (i.e., Rishbeth et al., 1969; Rishbeth et al., 1987; Rishbeth and Müller-Wodarg, 1999 and references therein and subsequently). However, this description neglects horizontal advection of height gradients on a constant pressure surface, i.e. $\vec{V}_h \cdot \vec{\nabla}_p z$, evident in equation (12) and included in the lifting wind component, making it an incomplete description. This will prove important when these terms are assessed in Section 5.

The temperature tendency (i.e., local time derivative of temperature) in pressure coordinates may now be expressed by expanding the internal energy equation 7 in section 2 as,

$$\left(\frac{\partial T}{\partial t}\right)_p = -\left(\vec{V}_h \cdot \vec{\nabla}_p T + \omega \frac{\partial T}{\partial p}\right) + \frac{\omega}{\rho \bar{c}_p} + \frac{\dot{Q}^*}{\rho \bar{c}_p} \tag{16}$$

This form states that the local temporal change in temperature at its current position in pressure coordinates is dependent on 1) horizontal and vertical advective terms, 2) vertical winds associated with pressure changes following the motion, and 3) diabatic heating rates. Note that the use of $c_p$ accounts for enthalpic changes introduced by internal energy and *volumetric* changes of the material element irrespective of vertical winds – see section 2. Furthermore, vertical winds defined in pressure coordinates represent the contribution to local changes in the gas temperature due to vertical advection and the total derivative of pressure with time. Only a *portion* of the height-based vertical wind contributes to the energetics as shown by equation (14). The next section applies these general thermodynamic practices to the thermosphere using control volume analysis to understand the inner workings that involve vertical motions and their influence on the internal energy budget of the thermosphere.

## 4   Control Volume Analysis

To assist in the interpretation of the relations derived in the previous section regarding energy changes, a thermodynamic control volume analysis is applied to the upper thermosphere. A control volume (CV) is an identifiable volume of gas defining the material element where the boundaries are imaginary but helpful in defining a system for which to apply a thermodynamic analysis. The defined volume occupied by an unchanging mass of gas can be deformed or advected by fluid motion. In the following analysis, two thermodynamic conditions will be applied to the thermospheric gas: 1) a closed system where only energy can cross the CV boundary and 2) an open system where energy and mass can cross the CV boundary.

As described in section 2, vertical motion couples to the internal energy equation by its role in expansion/compression work processes. This is captured by the rate of enthalpic change caused by changes in the gas volume and/or pressure with time. A thermospheric gas parcel can experience volumetric change, without the influence of vertical motion, by volumetric expansion or compression work. This work process occurs under constant pressure and is accounted for in the internal energy equation by defining the specific heat of the gas at constant pressure, $c_p$. Specific heat at constant pressure relates the enthalpic change to a temperature change when undergoing an isobaric volumetric expansion or compression, see section 2.

A thermospheric gas parcel can also experience a change in pressure with time causing a change in internal energy due to the work required to expand or compress the parcel when experiencing differential pressure from its surroundings. Vertical motion in the thermosphere moves parcels across largely stratified pressure surfaces and thus induces differential pressure between the parcel and its surroundings. The parcel will expand or contract to offset this pressure differential by exchanging energy between work and the internal energy of the gas if the process is performed adiabatically, i.e. no heat transfer during the process. The general attribution of this process in the internal energy equation is captured by the adiabatic heating / cooling term. However, vertical motion responsible for this type of energy transfer may only be a *component* of the total measured vertical motion depending on the specification of the vertical coordinate frame that defines the vertical motion. These energy considerations will be elucidated by the following thermodynamic analyses of a closed and open thermosphere CV.

## 4.1 Closed System

Figure 1 presents a closed thermodynamic system analysis applied to a volume of thermospheric gas having undergone a change from state 1 to state 2. The system under analysis is defined by the dashed lines, and the imaginary boundaries of the CV are given by solid lines. The CV occupies a horizontal area defined by $dA$ and vertical dimensions in either pressure, $Dp$, or height, $Dz$. The upper boundary to the control volume in Figure 1 is represented by a frictionless piston head whose weight equates to the total weight of air above the constant pressure level, defined in Figure 1 as $p_1$. An increasing height range is defined on the right side of the CV while a decreasing pressure range is provided on the left side. Solid horizontal lines within the CV define lines of constant pressure. The separation of lines represents the change in pressure with altitude that balances with the changing gas density via hydrostatics. The pressure force across the piston area is in mechanical equilibrium with the weight of the air above, such that the piston is at rest. Total mass ($M_{cv}$) is fixed within the control volume and defined using either of the two vertical coordinates by $M_{cv} = \rho dADz$ or $M_{cv} = dADp/g$. Figure 1 defines state 1 where the thermospheric gas properties are in thermodynamic equilibrium. A process must occur to change the state of the gas and its associated properties. Some of those state-1 properties at $p_1$ are listed along with properties representative of the CV.

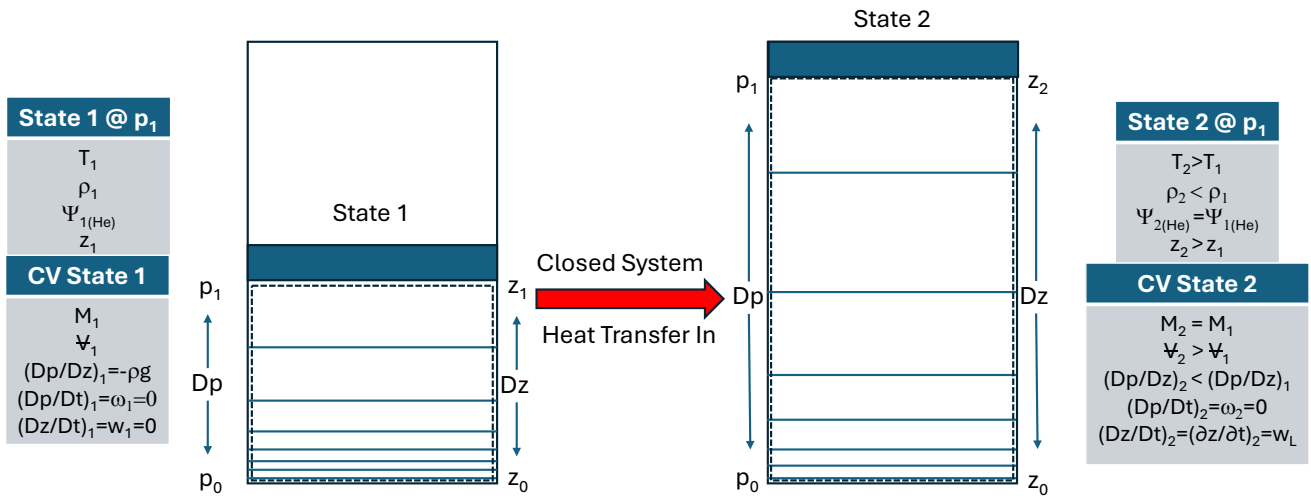

Figure 1. CV analysis for a closed system with heat transferred into the system causing a system change from state 1 to state 2. Thermospheric properties at pressure level $p_1$ (temperature ($T$), mass density ($\rho$), mass mixing ratio for helium ($\Psi_{He}$), and altitude ($z$)) and for the entire control volume (total mass of the gas ($M$), total volume ($\forall$), total pressure change with respect to height ($DP/Dz$), total pressure change with respect to time ($DP/Dt$), and total height change with respect to time ($Dz/Dt$)) are listed for each state.

A heat transfer process across the CV boundary has been directed into the gas. This causes a change in system properties, and, after adjustments, a balance occurs between energy transferred in and out of the volume resulting in a new equilibrium state defined by state 2 in Figure 1. During the adjustment process to this new state, temperature will increase due to heat transfer into the gas and the pressure at the base of the piston will initially rise. However, the incremental pressure differential on the piston head will cause the piston to rise until pressure re-equilibrates to its original state-1 value and the piston is once again in mechanical equilibrium. This is described as a quasi-equilibrium, isobaric process as the pressure adjustment back to $p_1$ is rapid. In fact, with hydrostatic equilibrium imposed, the pressure adjustment is instantaneous. This is precisely what happens in GCM's that universally impose hydrostatic equilibrium. Furthermore, this process is considered isobaric as $p_1$ has not changed throughout the expansion process (i.e., the gravitational force per unit area of the piston head has not changed).

A noted change in volume from state 1 to state 2 indicates volumetric expansion as the $P_1$ level increases in altitude. According to the first law, energy must be expended by the gas in doing expansion work to account for the change in volume, and this energy comes from the internal energy of the gas. Consequently, some of the internal energy gained in the heat transfer process by the gas will go into the work it takes to expand the volume rather than raise the temperature. Thus, for the same amount of heat transfer, the temperature of the gas does not increase as much as it would if the volume remained fixed. This is considered an enthalpic change to the system where, under a constant pressure process, the enthalpy of the gas will change to account for this volumetric expansion. This is accounted for in the energy equation by using the relationship that defines $c_p$, see section 2.

Note, in this scenario there is no gas circulation, simply an isobaric, volumetric expansion. Thus, on the $p_1$ pressure surface, there is no change in pressure with time, i.e. $Dp/Dt = \omega = 0$. However, in Cartesian coordinates, there is a change of the piston head position with time (between state 1 to state 2) with $z$ increasing and thus $Dz/Dt = w>0$. Using the expansion of the Lagrangian derivative in pressure coordinates and the lack of gas motion, only the local time derivative of $z$ contributes to $w$. Recall, this is defined as a portion of the lifting wind, $w_L$, and $\omega = w - w_L = 0$ which is consistent with $Dp/Dt=0$. Furthermore, it demonstrates that the lifting wind found in Cartesian coordinates on a constant pressure surface does not contribute to the energetics of the gas. To be complete, there is a change in potential energy of the system with the piston head increasing its position in height. This form of energy is accounted for by the mechanical energy equation and is not part of the internal energy equation.

In this new equilibrium state 2, there are several diagnostic equations we can apply to describe the state of thermospheric properties. These are the ideal gas law, hydrostatic balance, and diffusive equilibrium. Applying the ideal gas law to the pressure surface at the bottom of the piston for state 2, i.e., $p_1$, the temperature increase is balanced by a mass density decrease to keep pressure constant. As noted, the base of the piston has the same pressure in state 2 as state 1 but its height has increased. Accordingly, the vertical pressure gradient within the CV must decrease from state 1 to 2 and the total mass density must also decrease to maintain hydrostatic balance throughout the CV. This is consistent with the total volume increasing while the total mass remains fixed to cause the total mass density within the control volume to decrease.

Diffusive equilibrium accounts for the balance of a species' partial pressure with its own weight. For the example provided, a species mass mixing ratio – such as helium, $\Psi_{He}$, – will remain the same from state 1 to state 2 on a constant pressure surface, as all species will experience the same relative change in individual scale height due to the temperature rise. If viewed from a constant altitude within the CV, the pressure, temperature, and mass density all increase, while the helium mass mixing ratio will decrease. Thus, organizing properties on a constant pressure surface removes the effects of isobaric, volumetric expansion that can complicate the interpretation of property changes on a fixed altitude.

## 4.2  Open System

### 4.2.1  Heat transfer and Mass Transport

Figure 2 provides a thermodynamic analysis similar to Figure 1 but with an open system allowing both mass and energy to pass through the CV. Again, heat is transferred into the CV but in this case, mass is able to move in and out of the system due to gas circulation. State 3 is a new equilibrium state of the thermospheric gas where the heat transfer rate in and out of the system and the mass flow rate in and out balance, such that, the total mass in the CV is kept constant. The mass flow rate out of the system by the diverging winds at $p_1$ is balanced by the mass flow rate into the system by converging winds at lower levels. It should be noted that for this scenario the pressure in the region of converging winds must decrease somewhat to produce a pressure gradient that supports convergent motion. This is similar to a schematic provided in Smith (1998, figure 2) with the precise behavior dependent on where energy is injected into the volume and how energy is removed from the volume. Other properties within the control volume are at equilibrium; they may vary spatially but do not change with time.

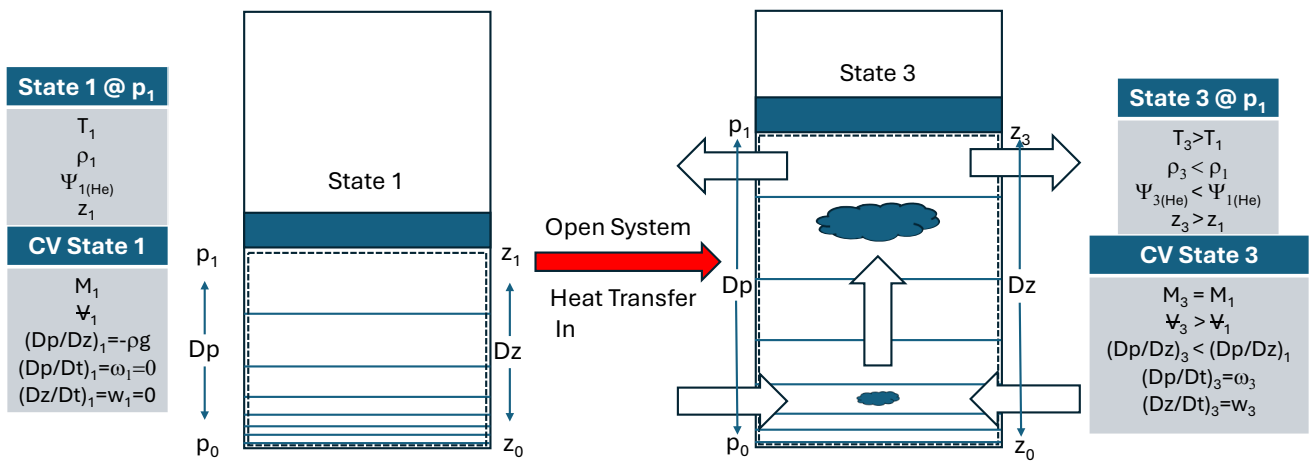

Figure 2. CV analysis for an open system with heat transferred into the system indicated by the red arrow. The white arrows and their direction indicate a balanced mass flow rate into and out of the system to keep the total mass of the CV fixed. State 3 represents a new equilibrium state once heat transfer and mass flow rate are balanced. As described in Figure 1, thermospheric properties at pressure level $p_1$ and for the entire control volume are defined for each state.

Once again, with heat transferred into the system at state 1, the gas volume expands and raises the height of the imaginary piston, thus enthalpy changes as before. However, this action is accompanied by gas circulation and, in particular, vertical winds in both pressure and height frames. Vertical winds transport mass from one level to another, defined as a parcel of gas, to satisfy mass continuity requirements. However, the gas parcel transports its lower-level properties to the upper-level region, if the motion is sufficiently rapid to not mix with its surroundings while being transported upward. Thus, upward transport of mass from the lower levels is at a higher pressure and the gas parcel adiabatically expands as it rises to meet the lower pressure in the upper levels. Now, the Lagrangian derivative of pressure, $Dp/Dt = \omega \neq 0$, and the energy associated with this adiabatic expansion must be accounted for in the energy equation, further altering the enthalpy of the gas at the $p_1$ surface. Adiabatic expansion of the gas parcel extracts from the internal energy of the gas and, therefore, induces cooling. Consequently, the temperature at $p_1$ for state 3 is still warmer than state 1 but cooler than state 2, given in Figure 1, due to this additional cooling contribution by pressure expansion. Consequently, the piston does not rise as high when comparing state 3 of Figure 2 to state 2 of Figure 1. Other properties are also impacted by this upward circulation, as listed in Figure 2. In particular, the mixing ratio of helium for state 3 will decrease at $p_1$ as compared to state 1 or 2. This is because the parcel mass from the lower level that is replacing those at $p_1$ have a lesser mass fraction of helium to the total mass within the gas parcel.

### 4.2.2 Mechanical Work and Mass Transport

A third scenario is depicted in Figure 3. Here, the system is open, but no heat is transferred into the system, i.e. adiabatic conditions. Instead, only mechanical work is performed on the gas at $p_1$ resulting in divergent flow. In the thermosphere, this can be caused by spatially varying momentum forcing in the upper region of the CV deforming the flow field, such as mechanical work performed on the gas by ion drag forces. This scenario induces 3-D circulation, schematically prescribed to be the same as Figure 2. However, without the accompanying heat transfer, the gas at $p_1$ cools solely by the induced upward vertical winds and adiabatic expansion, and the altitude of $p_1$ decreases. In this scenario, there is an upward vertical wind following the parcel motion but $w_L$ is of opposite sign. The component of height-based vertical wind contributing to the energy budget is then $w - w_L$, or $w_D$, and that equals $-\rho g w_D = \omega$. This type of circulation is often called an indirect circulation as it is not induced by direct heating but by mechanical work performed on the gas (e.g., Hsu et al., 2016). From an energy perspective, the process is adiabatic and reversible resulting in entropy being conserved. Approaches in the lower atmosphere have used entropy conservation (or potential temperature) to describe fluid flows but such special conditions do not persist in the thermosphere as diabatic processes can occur rapidly.

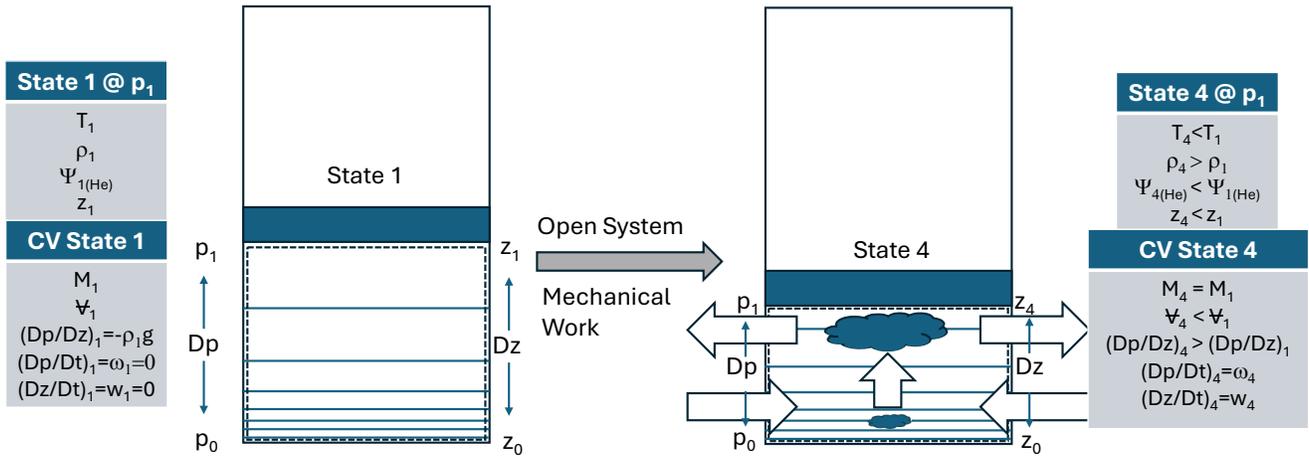

Figure 3. CV analysis for an open system with no heat transfer but mechanical work transferred into the system. The arrows and their direction indicate a balanced mass flow rate into and out of the system to keep the total mass of the CV fixed. As described in Figure 1, thermospheric properties at pressure level $p_1$ and for the entire control volume are defined for each state.

For this scenario, only mechanical work is done on the gas within the CV that causes the upper-level flow to diverge and the lower-level flow to converge, similar to the case for the open system with heat transfer. Mass flow rates remain balanced so that total mass within the CV remains constant, as in state 3, but properties are altered throughout the column due to the flow change. The first effect is to induce upward vertical motion in response to the upper-level divergent flow followed by lower-level convergent flow to maintain mass continuity, i.e., mass build up or depletion must be replaced by transferring mass between the two regions. The required upward vertical wind draws air into lower pressure which expands the gas resulting in adiabatic cooling and a temperature decrease. This decrease in temperature causes a pressure differential at the base of the piston which is relieved by moving the piston downward. This is a quasi-equilibrium isobaric process resulting in a volume decrease. In equilibrium, the pressure at the piston for state 4 is the same as state 1, while the state 4 temperature is decreased. Here, the volume of the CV decreases while the total mass remains constant resulting in an increase in the CV mass density. Assuming no heat transfer occurs across the boundary, this mechanical work process is accompanied by a reversible, adiabatic compression process such that entropy is conserved under this scenario. Thus, mechanical work processes do not change the entropy of the system while heat transfer processes do. Other properties of the system are similar to the heat transfer process described in Figure 2. This is a kind of circulation mixing that can alter species' mixing ratio on constant pressure surfaces. Thus, by determining the divergence of the horizontal wind at different pressure levels throughout the column, vertical and thermal responses can be estimated.

### 4.2.3 Lifting Wind and Horizontal Advection

One last scenario involves the horizontal advection of height surfaces on a constant pressure surface. In terms of a CV, the point under evaluation no longer has altitude and pressure surfaces aligned. Figure 4 illustrates the analysis on a constant pressure surface of the height advection term, $\vec{V}_h \cdot \vec{\nabla}_p z$, as adapted from Salby (1996). Height contours are drawn to indicate variations in height along the isobaric surface. The term is analyzed at the location of the unit vectors **i**, **j**, and **k** shown in the figure. Recall that this coordinate system is nonorthogonal such that the **k** unit vector is not normal to the pressure surface. However, **k** constitutes the vertical component of a right-handed coordinate system, such that a vertical wind is defined positive upwards in the **k** direction. The **i** and **j** unit vectors represent the direction of the zonal (positive east) and meridional winds (positive north), respectively, and for clarity the horizontal *x*-*y* plane is vertically offset in the figure to see the horizontal vector behavior. The horizontal wind in the advection term is, thus, the vector of the two wind components directed along the *x*-*y* plane. In the *p*-system, the gradient of height on a constant pressure surface is defined as

$$\vec{\nabla}_p z = \left(\frac{\partial z}{\partial x}\right)_{ypt} i + \left(\frac{\partial z}{\partial y}\right)_{xpt} j$$

where the subscripts indicate the variables being held fixed. Thus, the gradient lies in the horizontal plane capturing how height varies along the isobaric surface. The dot product operation to compute the advection term is then all performed in the horizontal plane, as indicated by the horizontal wind vector and height gradient vector in Figure 4. This manifests as a contribution to the vertical wind in the **k**-direction in a manner similar to a float bobbing in the water as surface waves move horizontally. This term adds to the lifting wind which also accounts for the height tendency on a constant pressure surface, see equation (12). These contributions to the lifting wind will be evaluated in the next section using the TIEGCM to determine how significant they are in describing height-based vertical motion on a constant pressure surface.

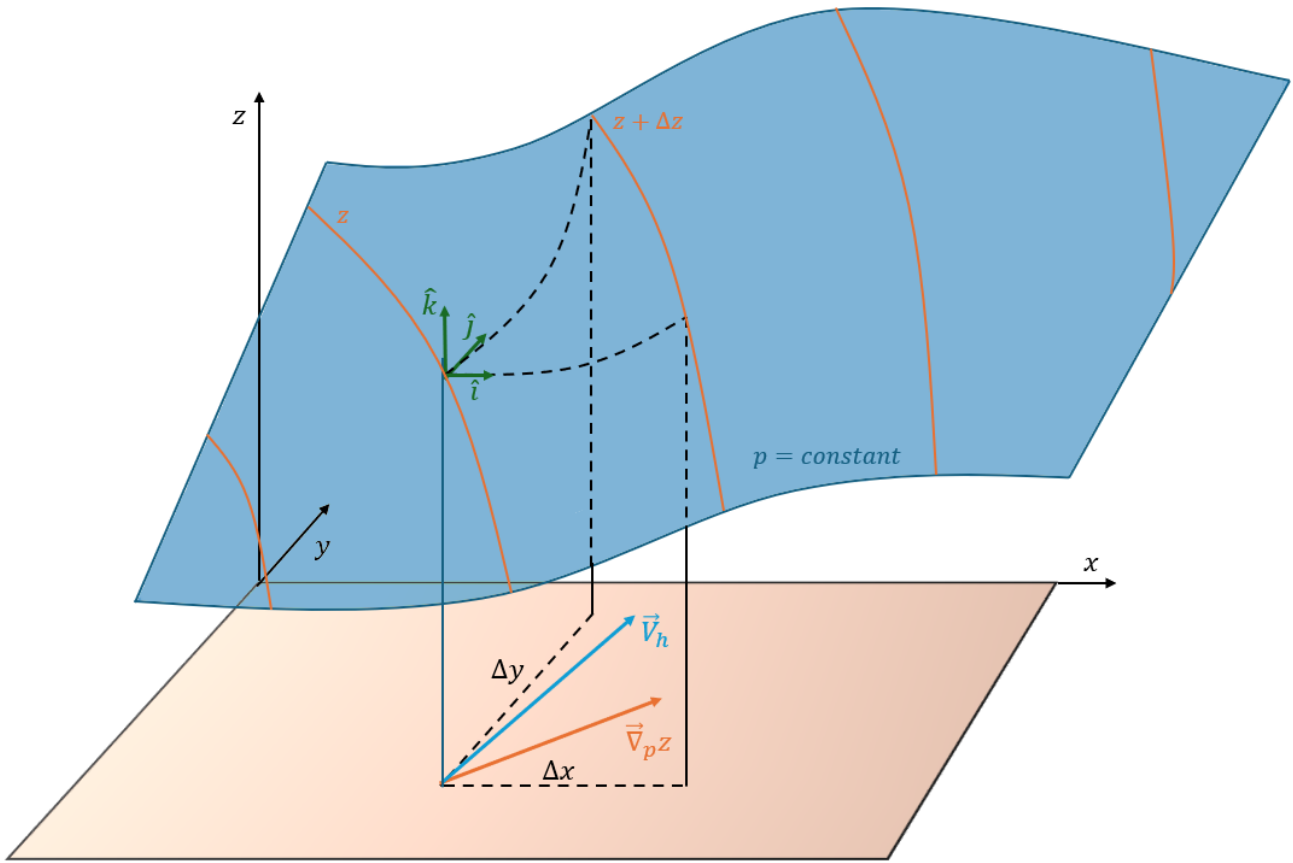

Figure 4. Height advection analysis on a constant pressure surface. The point under analysis is indicated by the location of the unit vectors **i**, **j**, and **k**. The constant pressure surface is free to undulate in this frame. Contours of height (orange) on the pressure surface are indicated to increase from left to right. The $x$-coordinate is longitude, and the $y$-coordinate is latitude. The orthogonal horizontal coordinate surface is vertically displaced from the pressure surface to allow for the horizontal wind and height gradient vectors to be distinguished. Adapted from Salby (1996).

# 5 TIEGCM Analysis

The CV analysis and scenarios described in section 4 provide insight into the interpretation of the inner workings that occur within the thermospheric gas and how they influence the energy budget. These physical processes are self-consistently captured by the TIEGCM by solving simultaneously the coupled mass, momentum, and internal energy equations for the thermospheric gas. The energy considerations associated with vertical winds are critically important for properly associating thermospheric property behavior under changing diabatic processes. The TIEGCM employs the dimensionless vertical log-pressure coordinate system, $Z = \ln(p_0/p)$, which introduces several non-intuitive modifications to the model equations that are described by Kasahara (1974) and summarized by Sutton et al. (2015) in their appendix. For this work, the hydrostatic, mass continuity, and internal energy equations in log-pressure coordinate system are most relevant. Also, the general use of height in previous sections is described in the model by its geopotential height which is a dependent variable in the four-dimensional space of the model, $\Phi = (x, y, Z, t)$.

The hydrostatic equation takes the form in the log-pressure system as,

$$\rho H_p = p_0 e^{-Z}/g \quad \text{where} \quad p = p_0 e^{-Z} \quad \text{and} \quad H_p = \frac{p}{\rho g} \tag{17}$$

Vertical winds in this coordinate system using the Lagrangian derivative are defined as,

$$W \equiv \frac{DZ}{Dt} \tag{18}$$

in units of [s$^{-1}$] with positive values directed upward in agreement with increasing Z with height. To convert this vertical wind into pressure coordinates, as defined in section 2, then $\omega = -pW$. Thus, it will be necessary in the energy equation of the TIEGCM to use this conversion to appropriately account for adiabatic heating or cooling.

As a hydrostatic model, the TIEGCM does not use the vertical momentum equation to derive vertical winds. However, this does not mean vertical winds cannot exist. It simply means that the vertical wind accelerations are generally much weaker than the accelerations caused by vertical pressure gradients and gravity. A method applied, at times, in meteorology to determine vertical winds is called the kinematic method (Holton & Hakim, 2013). This method vertically integrates the continuity equation to determine vertical winds on constant pressure surfaces. A caveat often associated with this technique in meteorology is the requirement that the wind field be somewhat ageostrophic to produce sufficient divergence to adequately estimate vertical winds. This can be a limitation when applying this technique in the lower atmosphere where the flow is largely geostrophic and vertical winds are weak. However, in the upper thermosphere, the flow is strongly ageostrophic (see Hsu et al., 2016) and vertical winds are of sufficient magnitude for this technique to be generally applied.

A further advantage in the use of log-pressure coordinates is that the continuity equation becomes an equation independent of time and mass density. The diagnostic continuity equation in log-pressure coordinates (see derivation in Sutton et al., 2015), is given as

$$\vec{\nabla}_Z \cdot \vec{V}_h + e^Z \frac{\partial \left(e^{-Z} W\right)}{\partial Z} = 0 \tag{19}$$

where the spatial gradient is defined on a constant log-pressure surface and $W \equiv DZ/Dt$ is the vertical velocity in the log-pressure coordinate system. As performed in the TIEGCM, the continuity equation is vertically integrated over log-pressure surfaces to determine the vertical wind, W, which can be used to estimate $\omega$. As discussed in section 2, the divergent component of the height-based vertical wind on a constant pressure surface, $w_D$ in units of m/s, is related to $\omega$ by the equation, $\omega = -\rho g w_D$, and can therefore be determined from the model output of vertical wind, W, by the expression,

$$w_D = \frac{pW}{\rho g} = W H_p \tag{20}$$

Furthermore, the terms contributing to the lifting wind, $w_L$, defined in section 2 can be evaluated on log-pressure surfaces to determine their relative magnitude and the relation of total vertical winds in the z-system, $w$, to the total vertical winds in the p-system, $\omega$.

The thermospheric energy equation solved within the TIEGCM is a variant of the internal energy equation provided in equation (16) with heat transfer and work terms defined specifically for the thermosphere system. The expression and description of terms can be found in past publications, such as Roble et al. (1988).

Here, we focus our energy analysis on the adiabatic expansion / compression term. As discussed in Section 2, the replacement of gas parcels on a particular pressure level with parcels transported from another pressure surface alters the local time rate of change in temperature. On a constant pressure surface, the adiabatic expansion / compression energy term, in units of K/s, is given as $\omega/\rho \bar{c}_p$ where the vertical wind in pressure coordinates is used. In log-pressure coordinates, the expression becomes

$$\dot{Q}_{adb} = -Wp/\rho \bar{c}_p \qquad (21)$$

If describing this term using a total vertical wind from the *z*-system on a constant pressure surface, only the divergent component of the height-based vertical wind is to be used and the expression for the adiabatic expansion / compression energy term in units of K/s is given as $-w_D g/\bar{c}_p$. As described in Section 3, either approach can be used with careful attribution when considering the height frame. The next section performs TIEGCM runs to evaluate vertical winds, their various components, and the adiabatic heating / cooling term.

## 5.1 Model Runs

The National Center for Atmospheric Research Thermosphere-Ionosphere Electrodynamics General Circulation Model (NCAR-TIEGCM, prior and henceforth referred to as TIEGCM V3.0) is used in this study with its improvements over V2.0 (Qian et al., 2014) described comprehensively by Wu et al. (2025). The model was run for 20 days to produce a diurnally reproducible state for June solstice conditions under moderate solar EUV fluxes (F10.7=150) and a Kp =1. A modest increase in geomagnetic activity (auroral hemispheric power of 55.56 GW, and a cross polar cap potential of 87.8 kV, i.e., Kp = 4) at 00 UT was then imposed for a 24-hour period. The horizontal resolution was set to 2.5º x 2.5º in terms of latitude and longitude, the vertical resolution was one-quarter scale height, and the time step was 5 minutes. The lower boundary condition was specified in terms of migrating and non-migrating diurnal and semi-diurnal tides, using the Global Scale Wave Model (GSWM). The following analysis focuses on model output after 18 hours of modest geomagnetic forcing.

## 5.2 Vertical Winds on Pressure Surfaces

Various forms of vertical winds derived from TIEGCM output on a normalized log-pressure surface are presented using the definitions provided in Section 3. Figure 5 illustrates vertical winds on a constant normalized log-pressure surface of Z=2.75 (~400 km) from the TIEGCM simulation converted to the height-based divergent vertical wind component, $w_D$, using equation (20) to provide values more generally familiar in units of m/s (remembering that this vertical wind component is proportional to the vertical wind determined on a constant pressure surface using the mass continuity equation – see equation (15)). Per model convention, a positive $w_D$ is upward. Figure 5 presents values of $w_D$ for the active period in the model run at 18 UT in the northern polar region. The model output shows enhanced divergent/convergent vertical winds during modest geomagnetic activity revealing a particularly structured and sustained alternating feature on the nightside. This vertical wind structure has upward (divergent horizontal winds) and downward (convergent horizontal winds) magnitudes, each exceeding 30 m/s. This component of the vertical winds in the $z$-system contributes directly to adiabatic heating / cooling rates via equation (21).

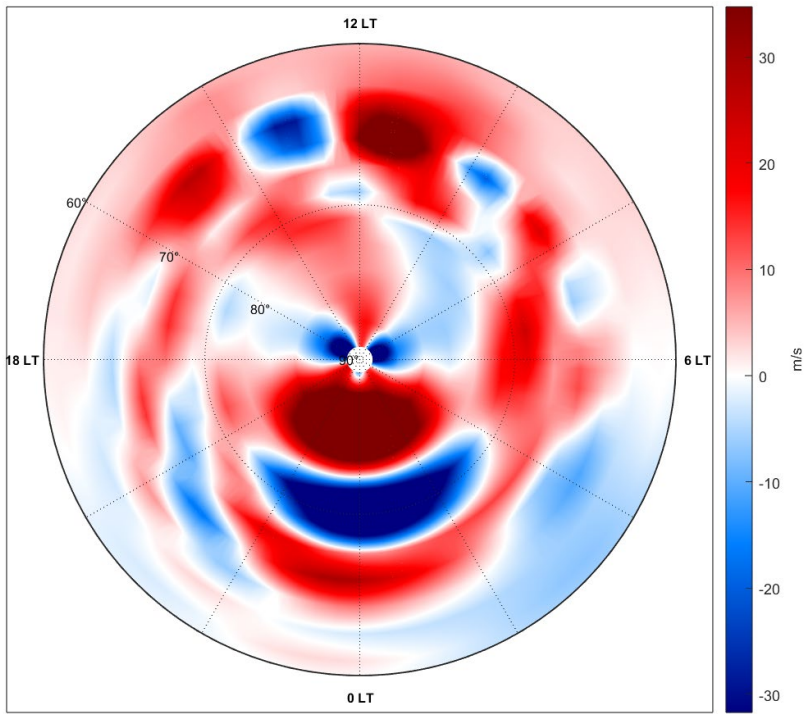

Figure 5. Northern hemisphere polar plot in local time and geographic latitude of the divergent vertical wind (m/s) determined from the TIEGCM simulation on the Figure 6 presents values of height-based lifting winds, $w_L$, in units of m/s, defined by the two terms in equation (12) computed using TIEGCM output. The lifting wind on a constant pressure surface is that component of the vertical wind in the z-system associated with the height tendency (local rate of change in height with time) and the horizontal advection of height, i.e. $w_L = \left(\frac{\partial z}{\partial t}\right)_P - \vec{V}_h \cdot \vec{\nabla}_p z$.

The lifting wind added to the divergent wind determines the total vertical wind in the z-system on a constant pressure surface. Plotted like the divergent vertical wind for the same time period in the simulation, Figure 6a illustrates the lifting wind determined from the model output. The magnitude of the lifting wind is appreciable and, for this time in the simulation, generally opposite in sign to the divergent vertical wind. Thus, at this point in the simulation and in these regions of enhanced divergent vertical winds, the height-based vertical wind in the z-system is appreciably less. The components contributing to the lifting wind on a constant pressure surface are provided in Figure 6 b, c. The dominant contributor to the lifting wind is that due to horizontal advection of height on a constant pressure surface. For this time in the simulation, the height tendency is small.

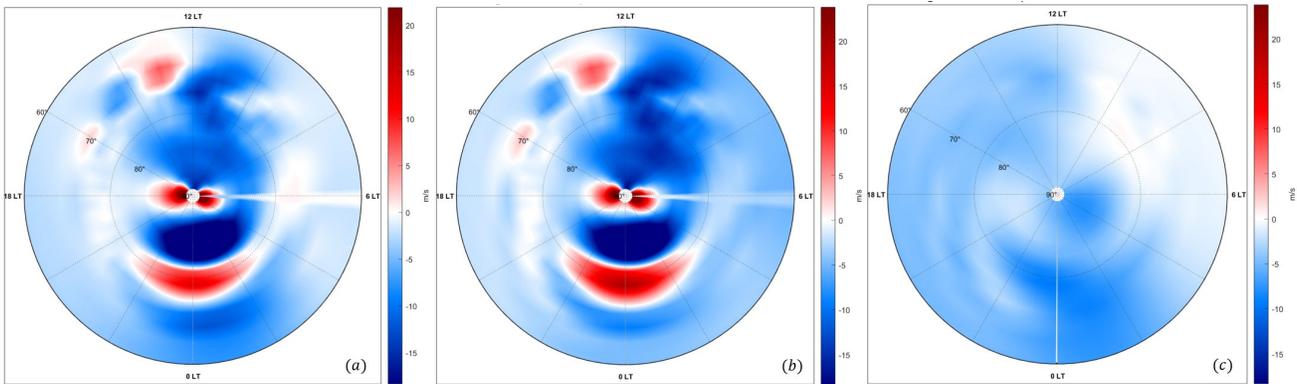

Figure 6. (a) Northern hemisphere polar plot in local time and geographic latitude of the lifting wind (m/s) and its components of (b) horizontal advection and (c) height tendency determined from the TIEGCM simulation on the Z=2.75 level at 18 UT.

Figure 7 plots the advection term values of the lifting wind in Figure 6b for a range of local times around midnight and extending in latitude from the pole to midlatitudes. This is analyzed in the manner described and presented in Figure 4, Section 3. The height of the constant pressure surface is particularly structured on the nightside with strong meridional gradients. Included on the constant pressure surface is the color contour of values for the lifting wind advection term illustrating where this term is significant. It is logical that the lifting wind would occur in regions similar to where divergent winds are present, see Figure 5. At this stage in the simulation, the model has reached a diurnally reproducible, albeit geomagnetically active, state. The height structure in Figure 7 is caused largely by adiabatic heating / cooling induced by the model's divergent vertical winds. Furthermore, horizontal winds associated with divergent / convergent motion are highly ageostrophic. Thus, height gradients will form in regions of ageostrophic winds resulting in significant height advection and opposite lifting winds relative to divergent vertical winds. Whether this is actually the case needs to be verified by observations but it demonstrates that the interpretation of vertical winds in the *z*-system on pressure surfaces comprises of several components which can be observationally challenging to resolve – see discussion in section 6 – and contribute differently to the energetics.

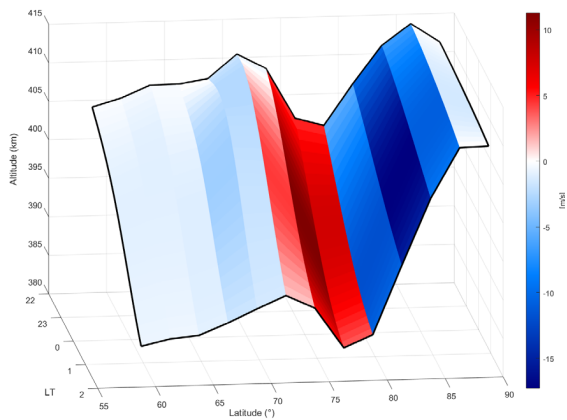

Figure 7. Northern hemisphere 3D plot in local time, geographic latitude, and altitude of geopotential height variations on the Z=2.75 pressure level at 18 UT from the TIEGCM simulation.

Dickinson and Geisler (1968) included this form of the lifting wind in their description of vertical winds in the *z*-system on a constant pressure surface but made the assumption, taken from Dickinson et al. (1968), that the advective term was small based on a scale analysis. This led Rishbeth et al. (1969, and many future references) to purport that vertical winds in the *z*-system on a constant pressure surface in the thermosphere consists of a barometric vertical wind, represented by the height tendency on a constant pressure surface, and a divergent vertical wind component, driven by horizontal winds on a constant pressure surface converging or diverging to maintain mass continuity.

In the lower atmosphere, the advective term is ignored because horizontal winds are largely geostrophic and such a wind on a constant pressure surface flows parallel to height contours, or perpendicular to the horizontal height gradient (Holton & Hakim, 2013). However, the upper thermosphere is highly ageostrophic (Hsu et al., 2016) and thus a significant component of the wind will flow along the horizontal height gradient making the advective term a significant contributor to the lifting wind. No past systematic study has evaluated the advective term. Our results indicate that the advection of height gradients on a constant pressure surface contributes significantly to the height-based vertical wind in the *z*-system but has a non-energetic influence on the gas. Appendix A provides an equivalent evaluation when transforming vertical winds from the *p*-system to the *z*-system, possible by the invertibility condition established by the non-zero Jacobian between the two coordinates.

# 6 Discussion and Conclusions

## 6.1 Adiabatic Heating and Cooling

Elucidating vertical wind descriptions in the thermosphere has important implications on evaluating its impact on the energy budget through its direct relation to adiabatic heating / cooling. Using the TIEGCM run presented in Section 5, the adiabatic heating / cooling rate in the upper thermosphere can be computed from the divergent vertical wind determined in the model via equation (21). Figure 8 plots the adiabatic heating rate in units of [K/s] on the log-pressure surface of Z=2.75 (~400 km height) along with percentage perturbations in temperature and mass density for the same time as presented in Figure 5 of Section 5.2. The percentage perturbations of temperature and mass density are determined by dividing their total fields by their respective low pass filtered background fields at this snapshot in time at 18 UT. The residual percentage perturbations represent high pass features whose scale sizes range from 500 km to a few thousand kilometers.

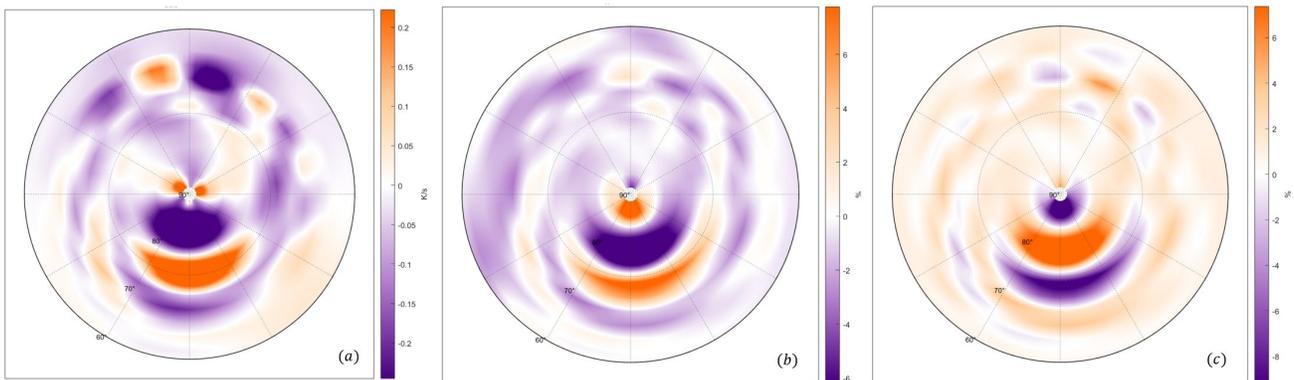

Figure 8. Northern hemisphere polar plot in local time and geographic latitude of (a) adiabatic heating rate [K/s], (b) perturbation temperature [%], and (c) perturbation mass density [%] determined from the TIEGCM simulation on the Z=2.75 level (~400 km height) at 18 UT.

The model run indicates that the vertical wind structure in the midnight sector is sustained for several hours resulting in significant adiabatic heating and cooling of the region resulting in temperature perturbations over several hundred degrees. The residual percentage perturbations in temperature exhibit strong spatial correlation with adiabatic heating / cooling. The nightside feature in adiabatic heating / cooling is clearly correlated with the neutral temperature perturbations producing ±6% change in regions of strongest heating and cooling about the background field. The temperature is influenced by all terms involved in the energy equation and a full energy analysis is required to assess the full temperature response. However, the adiabatic heating / cooling structure is clearly impressed on the temperature field as illustrated by the perturbation analysis.

The perturbation mass density in this nightside region on the constant pressure surface also exhibits similar structure to the adiabatic heating / cooling rates, and is generally anticorrelated with temperature to ensure pressure remains constant. On a constant pressure surface, vertical winds can also cause composition changes that alter the mean molecular weight of the gas. This contribution must also be considered when evaluating mass density changes on constant pressure surfaces (see Lei et al., 2010). However, the direct spatial correlation of density structures with temperature structures indicates that, at this stage of the simulation, the adiabatic heating/cooling process is strongly influential on these properties .

Clearly divergent vertical winds directly affect adiabatic heating rates which, in turn, significantly influence the energy equation and the structure in thermospheric properties. Thus, it is of high importance that the 3-D neutral circulation in the model be accurate. Presently, vertical and horizontal wind observations are limited as these 3-D structures span in range from a few hundred kilometers to thousands of kilometers. Observations must be distributed within and over this range of scale sizes to properly capture the kind of structures reproduced by the model. Adiabatic heating and cooling and the 3-D neutral circulation are inner workings of the I-T system that are critically important to the thermospheric energy, composition, and momentum budget that influence the final response in gas properties.

The outcomes presented in this Section and in Section 5 are produced by the TIEGCM without validation from observations. The self-consistency in the model between neutral momentum, energy, and continuity gives confidence in the vertical wind analysis. However, validation is required as there can certainly be observing scenarios where the relative magnitudes of these model terms differ – such as during wave propagation when the height tendency may dominate over advection of height gradients. However, the presented analysis demonstrates the issue that vertical wind observations and subsequent analysis/interpretation need to be very careful by defining appropriate coordinate frames and evaluating terms relative to that frame. Appendix A provides an equivalent evaluation of vertical winds and property responses when transforming from pressure to height as the vertical coordinate.

## 6.2 FPI Observations and Vertical Winds

A case in point is to apply this understanding to the interpretation of vertical winds retrieved from ground-based airglow emissions. Studies have been carried out to evaluate simulated *horizontal* thermospheric winds using ground-based 630.0nm Fabry-Perot interferometer (FPI) Doppler wind observations achieving reasonable success (e.g., Wang et al., 2021; Jiang et al., 2018 and references therein). Vertical winds from FPI 630.0nm emission observations have been more difficult to reconcile with theory with many observations reporting a broad range of vertical wind magnitudes (see Harding et al., 2017 and references therein). Makela et al. (2014) and Harding et al. (2017) attributed some of the discrepancy to observing artifacts. However, there is a fundamental premise to the interpretation of vertical wind measurements that involve vertical reference frames and observational frames that need to be reconciled.

Let's represent the 630.0nm airglow layer as a surface defined by its peak in volume emission rate (VER) being observed by a ground-based FPI. A general rule of thumb often applied to the 630.0nmm VER is that it occurs about a scale height below the F$_2$ ionospheric peak near an altitude of about 250 km. This assumption is generally due to the emission arising largely from O$_2$+ dissociative recombination. However, there is no basis that this 630.0 nm VER surface is fully aligned with height or pressure nor that horizontal gradients are nonexistent. In essence, the 630.0 nm VER surface represents a natural or generalized vertical coordinate system. To focus on vertical behavior relative to pressure (of interest for energy purposes as established in previous sections), assume the generalized coordinate system of the airglow surface is defined by the generalized vertical coordinate, *s=s(x, y, p, t)*, where it is a function of time, and the independent variables are horizontal spherical coordinates *x, y*, and vertical coordinate pressure. Assume there is a single-valued monotonic relationship between vertical coordinate *s* and pressure, when *x, y*, and *t* are held fixed (note: this assumption means that for every input value of *p* there is only one possible output for *s* and as that input *p*-variable changes, the *s*-variable either always decreases, increases, or is constant. Hydrostatic law is an example of a single-value monotonic relationship used when transforming between height and pressure in Section 3). Then, any scalar function in four-dimensional space may be expressed in either of two ways depending upon whether *s* or *p* is chosen as the vertical independent variable, and the scalar function may be transformed between the two vertical coordinate frames using the monotonic relationship.

As discussed in Section 3, defining the vertical coordinate in pressure is most desirable when evaluating the energy equation of the thermosphere because a measure of vertical wind in that coordinate system becomes, $Dp/Dt \equiv \omega$, which is directly related to adiabatic heating and cooling. However, the nature of the FPI observation places the measurement of vertical wind in the *s*-system, such that, $Ds/Dt \equiv w_s$. To relate the measurement of vertical wind in the *s*-system to a vertical wind in pressure coordinates, a set of relational equations are required. Section 3, and more detail in Kasahara (1974), provides the mathematical construct for describing a generalized vertical coordinate system for the transport equations of the atmosphere and this approach will be applied here.

From Kasahara (1974) and the discussion in Section 3, the Lagrangian derivative of a generalized vertical coordinate, defined above as $s$, on a constant pressure surface is,

$$w_s = \frac{Ds}{Dt} = \left(\frac{\partial s}{\partial t}\right)_p + \vec{V}_h \cdot \vec{\nabla}_p s + \omega \frac{\partial s}{\partial p} \tag{22}$$

where $w_s$ is the vertical wind in the $s$-coordinate system (FPI observing frame) and $\omega$ is the vertical wind in the pressure coordinate system, as described previously. Solving for the vertical wind in pressure coordinates results in the following expression,

$$\omega = \frac{\partial p}{\partial s}\left(w_s - \left(\frac{\partial s}{\partial t}\right)_p - \vec{V}_h \cdot \vec{\nabla}_p s\right) \tag{23}$$

This is very similar to equation (11) presented in Section 3 when describing the relation between height and pressure. In fact, if the vertical coordinate system of the airglow layer is aligned with height then these two equations would be exact and the vertical wind in the $z$-system would equal the airglow vertical wind. Furthermore, the lifting winds presented in the TIEGCM results in Section 5 would be representative of the components of vertical wind observed by the FPI that are not thermodynamically significant. In other words, the FPI measurement of vertical wind includes a lifting wind component if not aligned with pressure surfaces.

In general, it cannot be assumed that the airglow surface aligns with height or pressure surfaces. Figure 9 provides a general description expanding on the advection term description provided in Figure 4. The transformation is to convert a vertical wind observed in the *s*-system, $w_s$, to a vertical wind in the *p*-system. Equation (23) describes the transformation mathematically in the manner derived in Section 3. The point being analyzed is where the **i**, **j**, and **k** unit vectors meet. If the airglow layer is structured such that it is either changing with time or has horizontal gradients that are being horizontally advected on the constant pressure surface, the recorded Doppler shift in the vertical direction will include a lifting wind component in addition to any divergent wind component. Figure 9 indicates gradients of the airglow surface on the pressure surface. These are being advected by the existing horizontal winds displayed on the *x-y* plane. This must be subtracted from the observed vertical wind in the *s*-system. If there is a local time rate of change of the airglow surface at constant pressure, as indicated by the red circle in Figure 9, then that also must be subtracted from the observed vertical wind in the *s*-system. The resultant wind is then the divergent vertical wind in the *s*-system and is proportional to the vertical wind in the *p*-system by some unknown *s*-system to

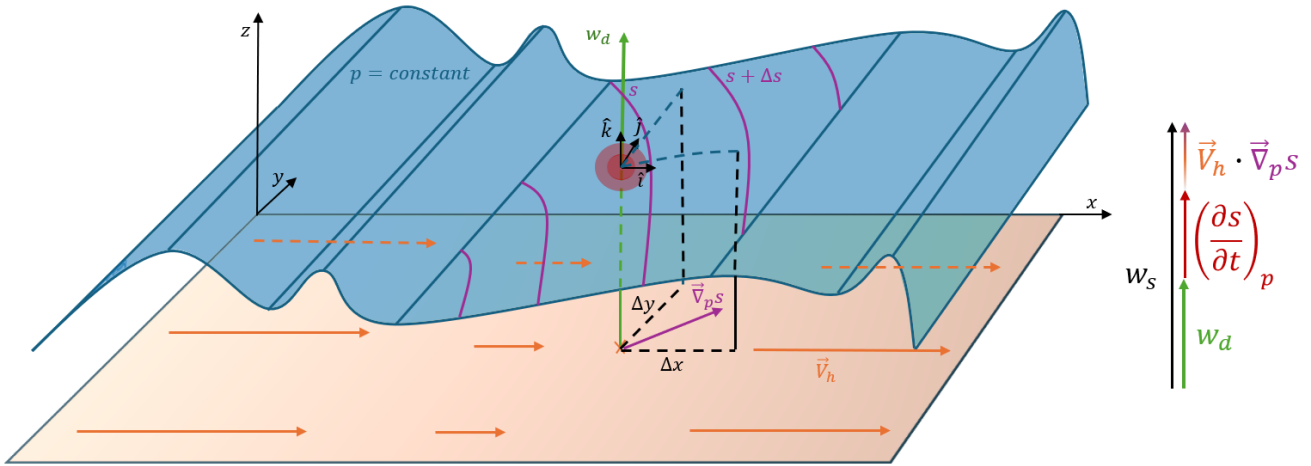

Figure 9. A general description of transforming vertical winds observed in the *s*-system to vertical winds in the *p*-system. The notation is the same as in Figure 4 when describing the advective term in the *s*-system. Added is a prescribed positive divergent vertical wind in the *s*-system and an expanding red dot to indicate time variations in the *s*-system that would contribute to the *s*-system total vertical wind. Refer to equation (23) to relate vector components shown on the right.

*p*-system functional relationship.

To define the FPI vertical winds on a constant pressure surface, requires determining the monotonic relation between pressure surface and the airglow surface, i.e. $\partial p/\partial s$, and how large a lifting wind component is present. The TIEGCM provides some evidence that the lifting wind can be a large percentage of the total vertical wind if the airglow surface is more aligned with altitude surfaces. If the airglow surface is more aligned with pressure surfaces then the FPI observations will include a significant divergent vertical wind contribution. Likely the airglow surface aligns with neither and can vary depending on environmental conditions and volume emission rate processes. Thus, an observation of vertical wind by a ground-based FPI observing an airglow layer is not equal to either the vertical wind on a pressure surface, as provided by the TIEGCM, or the vertical wind on a constant altitude surface, as provided by a model such as GITM (e.g., Yigit and Ridley, 2011), unless that airglow layer is aligned with the system's vertical coordinate system. The magnitude of the descrepancy is dependent on the horizontal structure of the volume emission rate's surface which can be difficult to ascertain from the FPI measurement alone. This can also impact the horizontal wind estimates from the FPI because a component of vertical motion will be in the line-of-sight measurement. This would be an issue in regions where large lifting winds are present.

## 6.3 Conclusions

This investigation has applied a generalized vertical coordinate system to clarify the transformations between vertical motions described in pressure, height, and volume emission coordinates. Simulations using the TIEGCM V3.0 model have been instrumental in revealing key differences and identifying previously overlooked components of vertical motion. Several important insights emerged from this analysis:

- A single, monotonic relationship between coordinate surfaces is essential for vertical motion to be consistently translated across coordinate systems.
- Height-based vertical winds defined on constant pressure surfaces comprise two distinct components:

- - A lifting wind which the TIEGCM simulation revealed to have significant magnitudes due to the horizontal advection of height on a constant pressure surface. This aspect has been neglected for over five decades. Well-designed observations will be required to resolve this contribution.
  - A divergent vertical wind which must be isolated from the total vertical wind in the height-based system either by subtracting the lifting wind component from the total vertical wind, or by deriving it from horizontal divergence estimates.
- Interchangeably, pressure-based vertical winds on constant height surfaces also consist of lifting and divergent components, influenced by pressure tendencies and horizontal advection across pressure gradients. (See Appendix A for further discussion.)

Through thermodynamic control volume analysis, the implications of these findings for thermospheric energetics are made clear:

- Accurate representation of vertical winds is critical for assessing adiabatic heating and cooling in the thermosphere.
  - Vertical winds determined in pressure coordinates directly affect internal energy via a linear relation to adiabatic heating and cooling.
  - Only the divergent component of the total vertical wind in the *z*-system contributes to thermosphere energetics.
  - Although the lifting wind component of the height-based vertical wind on a constant pressure surface can be substantial, it does not contribute to internal energy changes. The simulation showed that the lifting and divergent vertical wind components are often colocated and can be of opposite sign, complicating observational separation.

Model assessments under Kp = 4 geomagnetic conditions reveal that adiabatic heating/cooling can induce spatial perturbations in temperature, height, and mass density of significance and compete with direct energy inputs. Additionally, mass density variations observed at fixed altitudes (see Appendix A) can relate to divergent and lifting wind components through their association with temperature, pressure, and composition changes.

Finally, the generalized vertical coordinate approach highlights the challenges in interpreting vertical winds from airglow emissions, offering a potential resolution to the longstanding paradox between FPI-derived vertical winds and theoretical expectations. To disentangle lifting and divergent vertical wind contributions from FPI observations of the 630.0 nm emission, it is necessary to fully characterize the advection of horizontal emission structures and determine whether a monotonic relationship exists between the emission surface and pressure or altitude surfaces. Simulations indicate that strong lifting vertical winds often coincide with strong divergent vertical winds, further complicating efforts to separate energetic and non-energetic wind components in observational data.

# 7 Appendix: Vertical Winds on Height Surfaces

If the vertical wind transformation is from pressure coordinates to height coordinates, then a similar analysis to Sections 5 and 6 can be performed but evaluated on a fixed height surface. For this analysis, a constant height of 400 km will be used. The same TIEGCM run defined in Section 5.1 is used with model output interpolated from pressure surfaces closest to 400 km. The generalized form for the Lagrangian derivative of property $s(x, y, z, t)$ on a constant height surface is,

$$w_s = \frac{Ds}{Dt} = \left(\frac{\partial s}{\partial t}\right)_z + \vec{V}_h \cdot \vec{\nabla}_z s + w\frac{\partial s}{\partial z} \qquad (A.1)$$

If the property, $s$, is pressure, then the height-based vertical wind, $w=Dz/Dt$, relates to the vertical wind on constant pressure surfaces, $\omega=Dp/Dt$, by

$$w = -\frac{\omega}{\rho g} + \frac{1}{\rho g}\left\{\left(\frac{\partial p}{\partial t}\right)_z + \vec{V}_h \cdot \vec{\nabla}_z p\right\} \qquad (A.2)$$

This shows once again that the vertical wind in the *z*-system is equal to the sum of a divergent vertical wind and lifting wind, i.e.,

$$w_D = -\frac{\omega}{\rho g} \quad \text{and} \quad w_L = \frac{1}{\rho g}\left\{\left(\frac{\partial p}{\partial t}\right)_z + \vec{V}_h \cdot \vec{\nabla}_z p\right\} \tag{A.3}$$

However, in this case, it is the variation in pressure with time and the horizontal advection of pressure gradients on a constant height surface that determines the lifting wind. Due to the invertible condition between these surfaces via the hydrostatic relation, the terms for the divergent vertical wind and the lifting wind in these height coordinates have the same behavior as the divergent and lifting vertical winds presented using the isobaric coordinate system in Figure 6 of Section 5.2.

The analysis of adiabatic heating rates and temperature perturbations are also very similar to those presented in Figure 6a and b as there is very little change in adiabatic heating rates and temperature with height or pressure in the upper thermosphere. The mass density and pressure surfaces are most susceptible to a change in vertical coordinates with Figure A.1 presenting these properties for a fixed height at 400 km. Here, the adiabatic heating effects remain very prominently impressed on these thermospheric properties.

On a fixed height surface, pressure, composition, and temperature changes all contribute to a mass density change, whereas only temperature and composition changes contribute to mass density change on a fixed

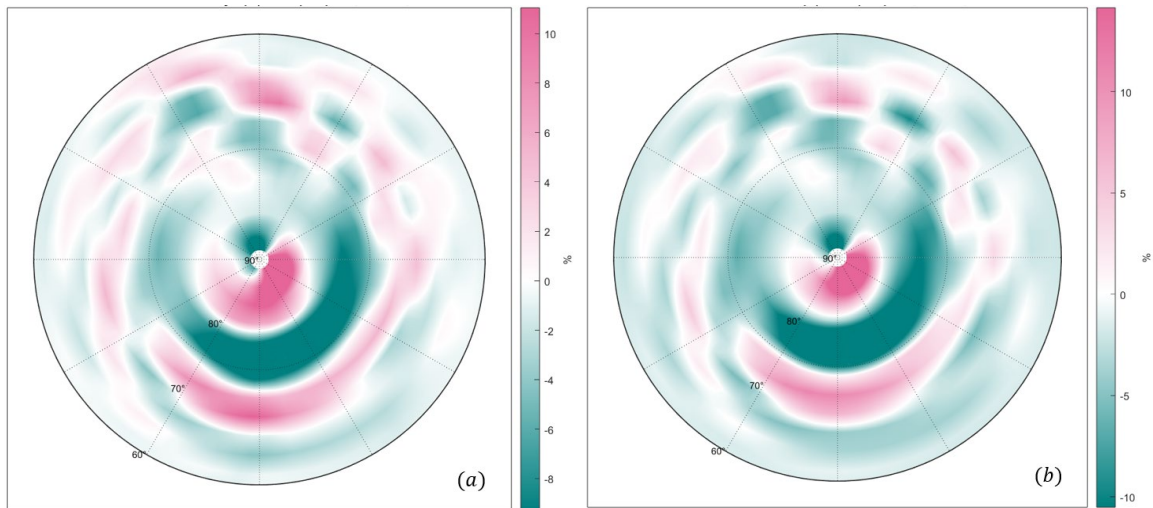

Figure A.1. Northern hemisphere polar plot in local time and geographic latitude of (a) perturbation pressure [%], and (b) perturbation mass density [%] determined from the TIEGCM simulation on *a constant height surface of 400 km* at 18 UT.

THAYER ET AL.: Energy Budget and Vertical Windspressure surface. As discussed by Lei et al. (2010), the response of mass density on a fixed height surface closely follows pressure variations. The mass density perturbation on a fixed height surface in Figure A.1 is nearly opposite to the mass density perturbations observed in pressure coordinates in Figure 8c. Given that the lifting wind component on a constant altitude surface depends on horizontal pressure gradients, the presence of lifting winds indicates pressure changes that strongly influence mass density. The lifting wind component is not the cause of the pressure changes but there is an association via equation A.3. The divergent wind component contributes to the adiabatic heating and cooling, impacting temperature and thus mass denisty. Furthermore, composition changes on a constant height surface are influenced by divergent vertical winds that alter the mean molecular weight of the gas and further complicate the mass density response. Thus, the ±8% mass density perturbations in Figure A.1 are attributed to both the divergent and lifting winds producing temperature, pressure, and composition changes on the fixed height surface.

The approach to computing the divergent vertical wind on a constant height surface should be similar to the approach used for a constant pressure surface, as long as hydrostatic balance is maintained. The general formula using the kinematic method on a constant pressure surface is derived from the continuity equation in the isobaric system, given as,

$$\left(\frac{\partial u}{\partial x} + \frac{\partial v}{\partial y}\right)_p + \frac{\partial \omega}{\partial p} = 0 \tag{A.4}$$

such that,

$$\omega(p) = \omega(p_T) - \int_{p_T}^{p} \left(\frac{\partial u}{\partial x} + \frac{\partial v}{\partial y}\right)_p dp = \omega(p_T) + (p_T - p)\left(\frac{\partial u}{\partial x} + \frac{\partial v}{\partial y}\right)_p \tag{A.5}$$

where $p_T$ represents the pressure at the top of the thermosphere. At the top of the thermosphere, vertical winds eventually go to zero, as most of the mass is kept within the model domain, and the pressure at the top of the atmosphere is much less than the pressure under evaluation. Consequently,

$$\omega(p) \approx -p\left(\frac{\partial u}{\partial x} + \frac{\partial v}{\partial y}\right)_p \tag{A.6}$$

Recalled from Section 3, the horizontal gradient on a constant pressure surface is defined as,

$$\vec{\nabla}_p = \left(\frac{\partial}{\partial x}\right)_{ypt} i + \left(\frac{\partial}{\partial y}\right)_{xpt} j$$

and the horizontal wind is defined as,

$$\vec{V}_h = ui + vj$$

The unit vectors **i** and **j** are in the zonal and meridional directions and are invariant to a change in the vertical coordinate. Thus, using the relation between the divergent vertical wind in height coordinates and the vertical wind in pressure coordinates, the expression solving for the divergent vertical winds from horizontal wind gradients at a constant height becomes,

$$w_D(z) = \frac{\rho(z_T)}{\rho(z)} w_D(z_T) - \frac{(p_T - p)}{\rho(z)g}\left(\frac{\partial u}{\partial x} + \frac{\partial v}{\partial y}\right) \tag{A.7}$$

Furthermore, the pressure and density at the top of the atmosphere is much smaller than the pressure level under investigation. Consequently,

$$w_D(z) \approx \frac{p}{\rho(z)g}\left(\frac{\partial u}{\partial x} + \frac{\partial v}{\partial y}\right) = H_p\left(\frac{\partial u}{\partial x} + \frac{\partial v}{\partial y}\right) \tag{A.8}$$

which once again equates to vertical wind in pressure coordinates via $w_D = -\omega/\rho g$, if the horizontal wind divergence is determined in either vertical coordinate. The expression using pressure scale height is similar to that discussed by Burnside et al. (1981), however the distinction made here is that this represents only a component of the total vertical wind in the *z*-system.

The divergent vertical wind is only a component of the total wind that would be determined by GITM or observed by an FPI assuming the 630.0 nm emission were aligned with height – see section 6. In fact, using the observed horizontal winds to compute the divergent vertical wind in this manner and comparing it with the measured total vertical wind from the FPI can give an indication of how much the lifting wind component contributes to the total vertical wind measured. In essence, the study by Anderson et al. (2011)

performed this exercise revealing differences between vertical winds derived from horizontal wind divergence estimates and those directly observed by the FPI in the vertical direction. Based on our fundamental description of vertical motion, that difference can be attributed to lifting winds which are part of the directly measured vertical winds.

**Acknowledgments.** This work was supported by NASA Geospace Dynamics Constellation contract 80GSFC22CA012. The manuscript benefited from helpful discussions with Dr. Eric Sutton and Anton Buynovskiy.

Jeffrey P. Thayer, University of Colorado, Aerospace Engineering Sciences Dept, Boulder, CO 80309 (e-mail: jeffrey.thayer@colorado.edu)